\documentstyle[preprint,aps,epsfig]{revtex}
\tightenlines
\begin{document}
\title{Equivalence of model space techniques    
 and the renormalization group
 for a separable model problem}
\author{S.K. Bogner and T.T.S. Kuo }
\address{ Department of Physics, SUNY, Stony Brook, New York 11794, USA\\}
\date{\today}
\maketitle
\begin{abstract}
Lee-Suzuki similarity transformations
and Krenciglowa-Kuo folded diagrams are two common methods used 
to derive energy independent model space effective interactions for nuclear many-body systems. 
We demonstrate that these two methods 
are equivalent to a Renormalization Group (RG) analysis 
of a well-studied problem in quantum mechanics.
The effective low-momentum potentials $V_{eff}$ obtained from model space methods are  
shown to obey the same scaling equation
for $V_{eff}$ that RG arguments predict.    
This  
indicates that model space methods might be 
of interest to those studying low-energy nuclear 
physics using Effective Field Theories (EFT).
We find the new result that all of the different energy independent model space techniques yield a unique low-momentum $V_{eff}$ when applied to 
the toy model under consideration.

\end{abstract}
\draft
\newpage
\section{Introduction}
There has been much effort over the past decade to describe 
low-energy nuclear phenomena  
using the techniques of Effective Field Theory (EFT) \cite
{Wein,ksw,orv,epel98,fsteele}. 
The main goal of an EFT description of these phenomena is not to out-perform
the traditional methods at fitting the experimental data.  Rather it is to 
provide a {\itshape model independent\/} effective low energy theory 
whose form is constrained by the 
symmetries of QCD, as well as to eliminate the uncontrolled
and unjustified approximations that are made in the traditional approaches.  
One of the more appealing features of an EFT treatment is the ability to 
reliably estimate the errors in calculations.  This allows   
one to consistently calculate observables to any chosen level of accuracy 
in principle. 
The success of 
various low energy effective theories (Landau Fermi Liquid theory,
the Fermi theory of Beta decay, the Standard Model, etc) results from the 
fundamental tenet of EFT:  {\itshape low energy observables are insensitive to 
the details 
of the high energy dynamics\/}.  Even if the 
high energy dynamics are unknown, one can still mimic their effects on the
low energy physics with local, model-independent operators that are 
consistent with the low-energy symmetries.  The corresponding 
coupling constants then implicitly contain the effects of the integrated out 
high energy degrees so that the low-energy physics of the full theory
is preserved.  EFT's are founded upon the ideas first
put forward by Wilson in his formulation of the "modern" 
renormalization group in the study of phase transitions.  
The early review article of Wilson and Kogut \cite{Wilson} and the more recent
papers of Lepage and Polchinski \cite{lepage,polch} illustrate these ideas 
with pedagological examples and a minimum of formalism. 

Shell-model theorists  
have long employed the notion of effective interactions that are 
defined only within a truncated model space, but that implicitly contain
the effects of the states that are being thrown away (i.e.- integrated
out) so that certain low energy observables 
are preserved from the full-theory 
\cite{ko90,krmku74,suzuki80,suzuki94,andre96,jensen}.
The qualitative similarities between the modern RG techniques 
and model space methods for deriving effective
interactions within a truncated model space are
obvious.
It would be interesting to study the {\itshape 
quantitative\/} similarities between the two approaches along the lines of the 
recent work of Haxton and Song \cite{haxton}, who seek to employ RG concepts
to formulate a "rigorous" shell model free of ad-hoc assumptions 
with the ability to reliably estimate errors. 
The scope of the current paper is a bit more modest and is similar 
in spirit to the 
earlier work of Fields et. al. \cite{fields}, where they  studied 
the effective interaction for a simple model problem    
by writing down a scaling equation for the 
"generalized G-matrix" as the UV regulator is 
varied and eventually sent to infinity. Our approach is 
more in line with the Wilsonian view of renormalization, 
as we keep the UV regulator 
finite throughout and instead study how $V_{eff}$ scales as we vary the 
boundary of the model space.    

The main
purpose of this paper is to study a separable model 
problem that allows closed form solutions of  
the Lee-Suzuki (LS) 
and the Krenciglowa-Kuo (KK) iteration schemes.  We find that
$V_{eff}$ derived from both schemes obeys the same scaling equation 
that one would obtain from a RG
analysis of the model problem \cite{Birse}.  Moreover,
we find the new and interesting result that {\itshape all\/} 
of the various 
model space techniques corresponding to different methods of solving
the decoupling equation 
(LS and KK are two specific schemes) give
the same low-momentum $V_{eff}$.  
In other words, the low-momentum $V_{eff}$ is unique and 
independent of the particular model space scheme one employs for the toy
model we consider. 
This is a suprising result as it is known that  
when one considers bound state (e.g- shell
model) problems the different model space techniques often converge to
different (i.e.- non unique) $V_{eff}$'s 
\cite{ko90,krmku74,suzuki80,suzuki94,andre96,jensen}.

\section{Model Space Techniques}  
  We turn now to a quick review of the methods one uses to derive energy 
independent
model space effective interactions.  Andreozzi has shown that both the Lee-Suzuki (LS)
and Kuo-Krenciglowa (KK) methods 
can be recast in the form of a general similarity transformation \cite
{andre96}.
In this formalism we  
perform a similarity transformation to obtain an effective 
Hamiltonian that acts only within the model space, but that 
preserves low-energy physics (spectra, scattering amplitudes, etc)
from the full problem.   
If we denote the 
projection operator onto the model space as $P$ and its complement as 
$Q$ (using the eigenkets of the unperturbed Hamiltonian as our basis) 
we have  

\begin {equation}
 S^{-1}HSS^{-1}\mid \Psi \rangle =ES^{-1} \mid \Psi \rangle 
\end {equation}

  We can write the above equation in an obvious way
by defining $H_{eff}\equiv S^{-1}HS$ and $\mid \chi \rangle  = S
^{-1}\mid \Psi \rangle $.  A convenient form of the similarity 
transformation is $S= 1 + \omega$, where $\omega$ is called the wave
operator and is defined to satisfy $\omega= Q\omega P$.  
It is easy to see that $S^{-1}= 1 - \omega$  and that the P-space
projections of the transformed and original states are the same, 
$P\mid \chi \rangle= P\mid \Psi \rangle$.  We want to exploit the
invariance of eigenvalues under similarity transformations to 
preserve a subset of the exact eigenvalues, but with  $H_{eff}$ {\itshape
acting only within the model space\/}.  In other words we want
$H_{eff}= PH_{eff}P$ and $\mid \chi\rangle = P\mid \Psi\rangle$ so that
the excluded Q-space completely decouples from the problem.  
A necessary and sufficient
condition for this is $QH_{eff}P= 0$.  Explicitly writing this out
gives the so-called decoupling equation for $\omega$

\begin {equation}
\omega PHQ \omega + \omega PHP - QHQ\omega - QHP = 0
\end {equation}
   
  The decoupling equation is a non-linear operator equation and different
methods of solution can give different answers.  Andreozzi has recently
shown that two iteration methods of solving it are equivalent to the Lee-Suzuki 
method and the Krenciglowa-Kuo folded diagram theory \cite{andre96}.  The
differences between the LS and the KK methods are most apparent when
one is dealing with bound state problems and the basis states are discrete
(e.g.- harmonic oscillator orbits).  If the model space is d-
dimensional, then the LS $H_{eff}$ reproduces the lowest d eigenvalues of
the full problem and the corresponding P-space projections of the exact
eigenstates.  Conversely, the KK $H_{eff}$ reproduces the d eigenvalues of
the full problem whose eigenstates have the largest P-space component (i.e.-
maximum overlap).  
The great benefit of Andreozzi's work is that it contains solutions of the
decoupling equation that are formally equivalent to the LS and KK schemes, 
but his methods are non-perturbative in nature and hence eliminate the need
to calculate irreducible vertex functions (i.e.-$\hat{Q}$-boxes)  and perform G-matrix resummations.   
Andreozzi's methods can furnish analytic results for simple toy models, and yet
they are extremely robust and easy to implement numerically for realistic bare
potentials (Paris, Bonn-A, V-18, etc...) \cite{Bogner}.
In deference to Andreozzi's simplification of KK and LS calculations, we 
hereafter refer to them as the Andreozzi-Krenciglowa-Kuo (AKK) and the 
Andreozzi-Lee-Suzuki (ALS) schemes.  We simply quote the relevant equations
and refer the interested reader to Andreozzi's paper for details.
\section *{The ALS Method}  
\noindent
Writing $\omega=\sum_{n=0}^{\infty}X_{n}$ and $\sigma_{n}=\sum_{m=0}^{n}X_{m}$, the ALS equations are

\begin {equation}
X_{0}=\frac{-1}{QHQ}QHP
\end {equation}

\begin {equation}
X_{n}=\frac{1}{q(\sigma_{n-1})}X_{n-1}p(\sigma_{n-1})\quad n=1,2,...
\end {equation}

\begin {equation}
p(\sigma_{n-1})=PHP + PHQ\sigma_{n-1}
\end {equation}

\begin {equation}
q(\sigma_{n-1})=QHQ - \sigma_{n-1}PHQ
\end {equation}

\noindent
Note that $p(\sigma_{n-1})$ and $q(\sigma_{n-1})$ are the P-space
and Q-space effective Hamiltonians at each step in the iteration, and
the iteration converges when $\sigma_{n}\approx  \sigma_{n-1}$. 

\section *{The AKK method} 
\noindent
Using the same notation as the ALS scheme,
the $X_{n}$ in the AKK scheme are given by the following equations,

\begin {equation}
X_{0}PHP-QHQX_{0} - QHP=0
\end {equation}

\begin {equation}
X_{n}p(\sigma_{n-1})-QHQX_{n}+\sigma_{n-1}PHQX_{n-1}=0 \quad n=1,2,...
\end {equation}

\section{Schematic Model with Separable Potential}
Both the ALS and AKK methods are easy to implement numerically for realistic
models of the bare $V_{NN}$ such as the Paris and Bonn-A potentials.  However, 
it would be nice to construct a toy model for which the ALS and AKK 
equations yield analytic solutions.  For this purpose, we consider the 
following schematic underlying (or full space) Hamiltonian  

\begin {equation}
H=H_{0} + g \mid \eta\rangle\langle\eta\mid
\end {equation}

\noindent
where

\begin {equation}
\langle i\mid H\mid j \rangle
= \epsilon_{i}\delta_{ij}
+ g\eta_{i}\eta_{j}
\end {equation}

\noindent
We will use the following formula many times in the analysis
that follows.
\begin {equation}
\langle i\mid H^{-1}\mid j \rangle   
= \frac{\delta_{ij}}{\epsilon_{i}}
- \frac{g\eta_{i}\eta_{j}}{\epsilon_{i}\epsilon_{j}}
\frac{1}{1 + gF}
\end {equation} 

\noindent
where 

\begin {equation}
F=\sum_{l}\frac{\eta^{2}_{l}}{\epsilon_{l}}
\end {equation}

\noindent
Note that we don't have to worry about the above expressions becoming 
singular ($\epsilon_{i}=0$), as we are always inverting 
in Q-space (where the $\epsilon_{i}\neq 0$).  

\section{ALS and AKK  Analysis of the Schematic Model}
For the remainder of this paper, all state 
labels refer to relative momenta, and it is assumed we are working with a 
particular partial wave (labels are suppressed).  
If we assume that the underlying potential
$V(k',k)\approx\ 0$ for $k,k' >\Lambda$ (either by a form factor or a sharp 
cut-off), then the truncated model space will consist of all states with 
$k\leq\frac{\Lambda}{s}$, where $s > 1$.  The excluded Q-space will then consist
of all states with momenta lying in the shell $\frac{\Lambda}{s}< k \leq \Lambda$. 
For a finite shrinking or decimation (e.g. $s=2$), we can use the formulas of the 
previous two sections to calculate the $X_{n}$ of both the ALS and AKK schemes analytically.
Unfortunately the expressions quickly become unwieldy for $n>1$, and it is not
at all obvious how one can sum all of the $X_{n}$ to obtain an analytic expression for the wave operator (and 
hence for $V_{eff}$).  The trick that gets us off the hook is to consider infinitesimal
decimations where $\frac{\Lambda}{s}=\Lambda - \delta\Lambda$.  In this way we are able
to calculate the wave operator analytically and obtain a scaling equation for
$V_{eff}$.  
Using the definition $V_{eff}=H_{eff}-H_{0}$, cavalierly ignoring factors of $2\pi$, and working in units where $\epsilon_{p}=p^{2}$, we have

\begin {equation}
V_{eff}(k',k)=V(k',k) + \int\limits_{\Lambda -\delta\Lambda}^{\Lambda}
V(k',q)\omega(q,k)
q^{2}dq
\qquad  k,k'
\leq \Lambda - \delta\Lambda
\end {equation} 

\noindent
For $\delta\Lambda\rightarrow 0$, this becomes a flow equation for $V_{eff}$

\begin {equation}
\frac{\partial}{\partial \Lambda}V_{eff}(k',k)=
-\Lambda^{2}V_{eff}(k',\Lambda)\omega(\Lambda,k)
\end {equation}

\noindent
Since we take the limit of $\delta\Lambda\rightarrow 0$, we can safely ignore
contributions to $\omega$ that are of $O(\delta\Lambda)$
and higher.
With this in mind, both the ALS and the AKK equations simplify considerably.
\section*{ALS for Infinitesimal Decimations}
Recalling that
each intermediate projection operator $Q=\int\limits_{\Lambda-\delta\Lambda}^{\Lambda}
q^{2}\mid q\rangle\langle q\mid dq\quad \sim O(\delta\Lambda)$, we can replace all
of the $p(\sigma_{n-1})=PHP + PHQ\sigma_{n-1}$ by $PHP$ in the ALS equations.  
It can be shown that at each step in the ALS iteration scheme we can write 

\begin {equation}
\langle q'\mid q(\sigma_{n-1}) \mid q\rangle=
\epsilon_{q}\delta_{q'q} + g\bar{\eta_{q'}}\eta_{q}
\end {equation}

\noindent
Where $\bar{\eta_{q'}}=\eta_{q'}(1+constants)$.
Consequently, we can use equations 11 and 12 to write 

\begin {equation}
\langle q'\mid q^{-1}(\sigma_{n-1}) \mid q\rangle=
\frac{\delta_{q'q}}{\epsilon_{q}} - \frac{g\bar{\eta_{q'}}\eta_{q}}{\epsilon_q'
\epsilon_{q}} + O(\delta\Lambda)
\end {equation}

\noindent
Naively, one might think that the $X_{n}$ are all infinitesimals of $O(\delta\Lambda)$
since the ALS equations all have an intermediate $Q$ projection operator. {\itshape This
is wrong\/}.  One must remember that by equation 16 , the operators 
$(QHQ)^{-1}$ and $q^{-1}(\sigma_{n-1})$ consist of a diagonal part proportional
to a $\delta$-function plus an off-diagonal part.  It is clear that when an 
intermediate infinitesimal projection operator $Q$ acts on the off-diagonal part it
results in a term of $O(\delta\Lambda)$ that can be dropped.  However, the $\delta$-function
portion survives the infinitesimal $Q$ projection and gives a finite contribution.
Therefore, we can make a second major simplification to the ALS equations by
replacing $(QHQ)^{-1}$ and $q^{-1}(\sigma_{n-1})$ by $(QH_{0}Q)^{-1}$.
The simplified ALS equations can be solved by addition, resulting in the following
linear integral equation for $\omega$

\begin {equation}
\omega=X_{0} + \frac{1}{QH_{0}Q}\omega PHP 
\end {equation}

\noindent
Rearranging terms, explicitly displaying indices, and abbreviating integrals
as sums,  we obtain

\begin {equation}
\sum_{p'}^{\Lambda-\delta\Lambda}\omega(\Lambda,p')A(p',p)=-\epsilon_{\Lambda}X_{0}(\Lambda,p)
\end {equation}

\noindent
Where we have defined the matrix $A(p',p)=(\epsilon_{p'}-\epsilon_{\Lambda})\delta_{p'p}
+ g\eta_{p'}\eta_{p}$.  Multiplying from the right by $A^{-1}$ we obtain

\begin {equation}
\omega(\Lambda,p)=-\epsilon_{\Lambda}\sum_{p'}X_{0}(\Lambda,p')A^{-1}(p',p)
\end {equation}

\noindent
Once again, we make use of equation 11 to obtain an explicit expression
for $A^{-1}$, viz,

\begin {equation}
A^{-1}(p',p)= \frac{\delta_{p'p}}{\epsilon_{p}-\epsilon_{\Lambda}}
-\frac{g\eta_{p'}\eta_{p}}{(\epsilon_{p}-\epsilon_{\Lambda})(\epsilon_{p'}
- \epsilon_{\Lambda})}\frac{1}{1+g\bar{F}}
\end {equation}
    
\noindent
where we have defined $\bar{F}$ by the following equation

\begin {equation}
\bar{F}=\int\limits_{0}^{\Lambda-\delta\Lambda}\frac{p^{2}\eta_{p}^{2}}
{p^{2}-\Lambda^{2}}dp
\end {equation}

\noindent
Since we eventually take $\delta\Lambda\rightarrow 0$, it is
clear that $\bar{F}\rightarrow \infty$ allowing us to ignore the second term in 
the above expression for $A^{-1}$. Noting that $X_{0}(\Lambda,p)=\frac{-g\eta_{\Lambda}
\eta_{p}}{\epsilon_{\Lambda}}$ and utilizing the simplified expression for $A^{-1}$
, we obtain our final expressions for the ALS wave operator $\omega(\Lambda,p)$
and the ALS flow equation 

\begin {equation}
\omega(\Lambda,p)=\frac{-g\eta_{\Lambda}\eta_{p}}{\Lambda^{2}-p^{2}} =
\frac{- V_{eff}(\Lambda,p)}{\Lambda^{2}-p^{2}}
\end {equation}

\noindent
and     

\begin {equation}
\frac{\partial}{\partial\Lambda}V_{eff}(p',p)=\frac{\Lambda^{2}V_{eff}(p',\Lambda)
V_{eff}(\Lambda,p)}{\Lambda^{2}-p^{2}}
\end {equation}

\section*{AKK for Infinitesimal Decimations}
For infinitesimal
model space reductions $\Lambda\rightarrow\Lambda-\delta\Lambda$, we simply
state the simplified AKK equations as we use the same arguments we used to 
simplify the ALS equations in the previous section.  Equation 7 simplifies to

\begin {equation}
X_{0}PHP - QH_{0}QX_{0}
=QHP
\end {equation}

\noindent
while equation 8 simplifies to

\begin {equation}
X_{n}PHP - QH_{0}QX_{n}=0
\quad n=1,2,...
\end {equation}

\noindent
It is easily shown 
that all of the $X_{n}$ are zero except for $X_{0}$.  {\itshape Hence, the AKK
wave operator is given by $X_{0}$ and it is found to be the same as the ALS
wave operator\/}.  
\indent
We can generalize the above results to cover {\itshape any\/} method of
solving the decoupling equation, by once again considering an 
infinitesimal $\Lambda\rightarrow\Lambda-\delta\Lambda$.  In this
case the non-linear decoupling equation simplifies to 

\begin {equation}
\omega PHP - QH_{0}Q \omega - QHP=0
\end {equation}  

\noindent
It is easily shown that the solution to this equation is the same as the ALS 
and 
AKK solutions. Hence, when  
we shrink the boundary of P-space by an infinitesimal amount, all of the 
various methods of solving the decoupling equation 
give the same equation for $\partial_{\Lambda}V_{eff}$.   
We shall find that we obtain the same scaling equation if we cut off all loop 
integrals in the Lippman-Schwinger 
equation and demand the observable scattering amplitude be independent of 
$\Lambda$; this is the RG method described in the next section.    
We can now invoke the semi-group  property of the RG 
and make a much stronger statement about our model space techniques.   
Even for {\itshape finite
\/} $\Lambda\rightarrow\frac{\Lambda}{s}$, the
different methods of solving the decoupling equation 
give a unique low-momentum
$V_{eff}$ for the toy problem under consideration.   
To the best of our knowledge,
this is a new and interesting result.  This result brings out yet another 
similarity
between model space techniques and the Wilsonian RG. In the application of
the Wilsonian RG to critical phenomena, there are many different methods of
integrating out the short distance physics to obtain an effective long-wavelength theory.  For example, there are infinitely many prescriptions for performing block
spin transformations and assigning values to the block variables for the Ising
model.  However, the differences between the various coarse graining procedures
are eventually washed out and one always converges to the same long-wavelength
effective theory.
\section{renormalization group analysis}
Since we claim that the model space techniques of nuclear many-body physics 
give the same scaling equation for $V_{eff}$ that a renormalization group 
analysis predicts, it is convenient to quickly review exactly what we mean by 
a "renormalization group analysis". In relativistic quantum field theory 
when one 
calculates physical observables perturbatively as a 
power series in the coupling $g$, one finds that the coefficients of the higher
order powers of $g$ are divergent.  These infinities are the result of loop 
integrals that diverge at high momentum, or equivalently at short distances.  
One can isolate these infinities by regulating the divergent integrals with 
a UV cut-off or any other convenient method (dimensional regularization, Pauli-
Villars, etc...), but then one is faced with the unpleasant result that  
calculated observables depend on the arbitrary regulator mass 
$\Lambda$ that has been introduced into the problem, as well as the particular
method of regularization.  As is well known, the way out of this dilemna is to
notice that for certain field theories (the so-called "renormalizable" theories)
we can add counterterms to the theory that have the same operator structure as
the terms present in the initial Lagrangian.  If we are sufficiently clever, we
can hide the regulated infinities in the coefficients of the 
counterterms so that when we calculate graphs using the new Feynman rules 
(that are induced by the counterterms), everything is finite.  Hence, one can absorb
the divergences by redefining the couplings and making them $\Lambda$-dependent
in such a way that the {\itshape physical observables\/} are $\Lambda$-independent and finite as $\Lambda\rightarrow\infty$.  Thus, a change in the UV cut-off
scale is compensated by a simultaneous change in the couplings and masses 
leaving the physics invariant.  This is the essence, albeit a grotesquely 
simplified description, of the renormalization group in QFT designed to
hide the infinities at short distances.  But by allowing $\Lambda\rightarrow
\infty$ we are assuming that the dynamics encoded in the original Lagrangian
holds true at arbitrarily large energies.  In the language of old fashioned 
perturbation theory, allowing loop momenta to go up to infinity corresponds
to summing over intermediate states of arbitrarily large energy and momentum.  But this presupposes that the intermediate states we are summing are devoid of
any "new" physics at arbitrarily small distances, which is clearly a bold 
assumption (supersymmetric partners?  quark and lepton substructure?  strings?)
since $1$ TeV $\neq\infty$!  About the only thing we can say for certain about
the high energy intermediate states is that they are highly virtual and thus 
localized in space-time.

   The modern approach to renormalization views $\Lambda$ as a 
physical scale at which unknown physics comes into play, and not as an 
artifact that is to be taken to infinity at the end of the calculation.  In
this way we don't sum over intermediate states we don't fully understand.  Yet
we can model the effects of these excluded states on the low energy physics by
recalling that their virtual nature means that they are propagated over very 
small distances and times (i.e.- they are localized).  We can include 
their effects on the low energy physics by writing down every possible local 
interaction that is consistent with low energy symmetries (model independence!), and we can parameterize our ignorance of the the excluded high energy states 
inside of the tunable couplings that multiply each local operator. 
Although there are an infinite number of local operators consistent with the
symmetry requirements, they all scale as inverse powers of $\Lambda$ 
allowing one to truncate the effective Lagrangian to any chosen level of 
accuracy.  The corresponding couplings can then be obtained from a finite 
number of independent experimental measurements, giving the theory predictive 
power to the chosen level of accuracy.  In this way the effective theory is
consistent with the underlying symmetries, the errors are reliably estimated, 
different quantities can be calculated consistently to the same level of 
accuracy, and one can infer the presence of "new physics" when the effective
theory fails.  

  We can employ similar ideas in analysing our toy potential model.
Here, we can impose a UV cut-off $\Lambda$ in the loop integrals of the 
Lippmann-Schwinger equation to reflect our ignorance of what is really going
on at short distances.  We then demand that $V_{eff}$ depend on $\Lambda$
in such a way that the low energy observables (i.e.- the scattering amplitudes)
are independent of $\Lambda$. 
For our separable toy model we can write the half-onshell (HOS) T-matrix as  

\begin {equation}
\frac{1}{\langle k'\mid T(\omega=k^{2})\mid k \rangle}
= \frac{1}{g\eta_{k'}\eta_{k}} 
-\frac{1}{\eta_{k'}\eta_{k}}\int\limits_{0}^{\Lambda}\frac{q^{2}\eta_{q}^{2}}
{k^{2}-q^{2}}dq
\end {equation}

\noindent
Since we want the physical quantities to be independent of this UV cut-off
$\Lambda$, we allow the coupling to "run" with $\Lambda$.  By setting
$\partial_{\Lambda}T^{-1}=0$ (and noting that all the $\Lambda$-dependence is 
inside $g$), we obtain a 
scaling equation for $V_{eff}$

\begin {equation}
\frac{\partial}{\partial \Lambda}V_{eff}(k',k)=
\frac{\Lambda^{2}V_{eff}(k',\Lambda) V_{eff}(\Lambda,k)}{\Lambda^{2}
- k^{2}}
\end {equation}

\noindent
This is same scaling equation for $V_{eff}$ that we derived using
model space methods.  We also note that this is the same equation obtained
by Birse et. al. \cite{Birse}, except that their $V_{eff}$ depends on the
energy as well (i.e. $V_{eff}(k',k)\rightarrow V_{eff}(k',k;E)$).  
We therefore conclude that model space techniques preserve the low energy HOS
T-matrix and  are equivalent to renormalization group methods when applied
to the separable model. 
\section{conclusions}
We have found that model space methods and renormalization group techniques 
give the same scaling equation for the low-momentum $V_{eff}$.  
This was accomplished by considering a separable potential model 
in which the model space techniques yield analytic expressions.
Moreover, we 
have found that all methods of solving the decoupling equation give the same
$V_{eff}$ for the toy model under consideration.  
To
the best of our knowledge, these are new and interesting results.  The
generalization of these results to realistic $V_{NN}$'s will appear in a 
forthcoming paper. 
\section*{acknowledgements}
We thank the INT at the University of
Washington for its hospitality and the Department of Energy for
partial support during the completion of this work.  

\end{document}